# AN EFFECTIVE TRANSFORMER-BASED CONTEXTUAL MODEL AND TEMPORAL GATE POOLING FOR SPEAKER IDENTIFICATION

*Harunori Kawano, Sota Shimizu*

Shibaura Institute of Technology, Japan

## ABSTRACT

Wav2vec2 has achieved success in applying Transformer architecture and self-supervised learning to speech recognition. Recently, these have come to be used not only for speech recognition but also for the entire speech processing. This paper introduces an effective end-to-end speaker identification model applied Transformer-based contextual model. We explored the relationship between the hyper-parameters and the performance in order to discern the structure of an effective model. Furthermore, we propose a pooling method, Temporal Gate Pooling, with powerful learning ability for speaker identification. We applied Conformer as encoder and BEST-RQ for pre-training and conducted an evaluation utilizing the speaker identification of VoxCeleb1. The proposed method has achieved an accuracy of 87.1% with 28.5M parameters, demonstrating comparable precision to wav2vec2 with 317.7M parameters. Code is available at

*https://github.com/HarunoriKawano/speaker-identification-with-tgp*

## 1. INTRODUCTION

With the advent of Transformer [1], deep learning has witnessed a breakthrough, leading to the proposition of numerous Transformer-based contextual models (TBCMs) [8, 18, 19] in the field of natural language processing. Furthermore, by employing self-supervised learning (SSL) [3, 20, 21, 22], it has become feasible to pre-train a TBCM, enabling learning with a limited amount of labeled data. Wav2vec2 [2] is an automatic speech recognition model composed of a feature encoder utilizing Convolutional Neural Network (CNN) and a contextual network employing Transformer. Similar to BERT's Masked Language Modeling (MLM) [3], it is able to adapt SSL by masking a certain proportion of input data in the contextual network.

In recent years, not only speech recognition but also speaker recognition has seen the application of pre-trained TBCMs [14, 23, 24]. The previous models used CNN, Time Delay Neural Network, and so on in the field of deep learning, and are specialized in speaker recognition [25, 26]. However, after the appearance of wav2vec2, it has become mainstream to use TBCMs that are able to apply various tasks. In [4], wav2vec2 is applied to speaker verification. It indicates that pre-trained TBCMs produce similar accuracy to conventional models. In [5], a large-scale TBCM named WavLM has been proposed. It has developed as a full-stack speech processing model and has achieved state-of-the-art performance on SUPERB benchmark [6] that contains some speech processing tasks.

Speaker recognition accuracy has been greatly improved by using pre-trained TBCMs. However, the sizes of the models and their computational complexity has also increased significantly. The enlargement of the model leads to an increase in required computational resources and a deficiency in real-time capability. Hence, utilizing more efficient models, rather than merely augmenting the sizes of the models, becomes a critically significant task.

In this study, we focus on speaker identification, the simplest task in speaker recognition, and propose effective end-to-end speaker identification model applied TBCM. We explored effective TBCM architecture by focusing on the hyper-parameters. Also, we propose a pooling method named **T**emporal **G**ate **P**ooling (TGP) designed for speaker identification. The main hyper-parameters that determine the structure of TBCM are hidden size and number of layers. It is known that increasing these parameters not only enlarges the size of the model but also enhances its performance. We estimated the appropriate parameters by comparing these two parameters and the performance of speaker identification. In order to verify the effectiveness of the proposed method, the methods are evaluated on speaker identification task of VoxCeleb1 [7].

## 2. METHODOLOGY

The proposed method consists of four elements: an encoder, a pre-training framework, a pooling layer, a classifier (Figure 1). A Conformer-based model [8] is used as the encoder. Conformer is a TBCM combined CNN and Self-attention and is designed for speech recognition. BEST-RQ framework [9] is used for the pre-training of speaker recognition tasks. BEST-RQ is designed to reduce the difference between pre-training and fine-tuning and is considered suitable for speaker recognition. A fully connected layer and AAM-Softmax [10]

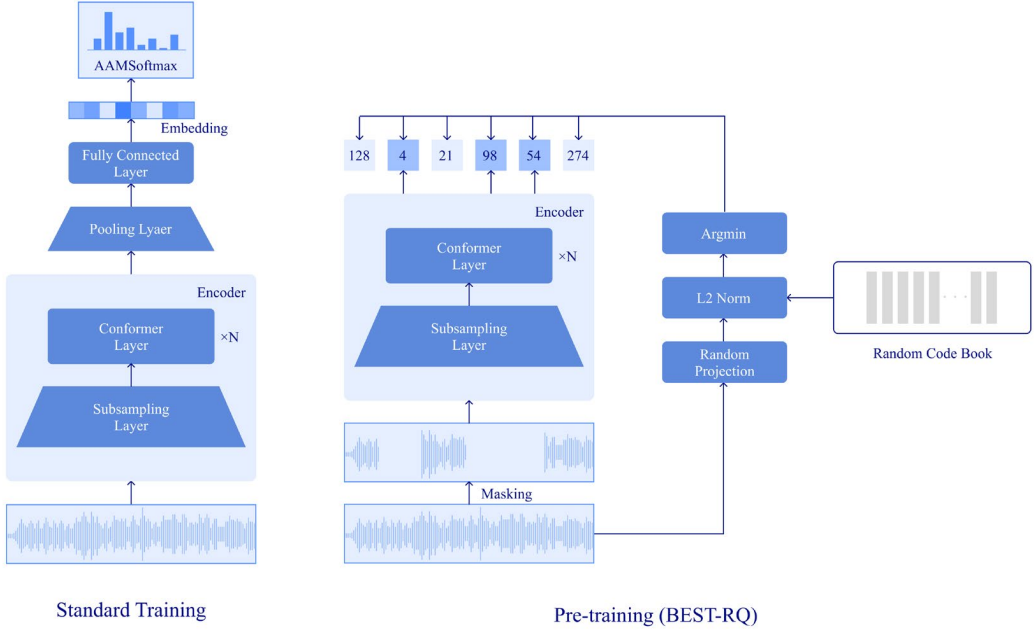

Figure 1: *Overview of the proposed method. The encoder is pre-trained with BEST-RQ framework before the standard training.*

with margin of 0.2 and scaling factor of 30 are used for the classifier.

### 2.1. Conformer-based Encoder

In the first step, inputs are passed through a subsampling layer. The subsampling layer consists of two CNNs with kernel size of 3 for each CNN, and a stride of 2. The input lengths are reduced to 1/4 through the process of subsampling layer. The activation function in the subsampling layer is used ReLU.

After the subsampling process, the hidden states are passed through a stack of Conformer layers. The Conformer layer consists of two feed forward modules, a multi-head self-attention module and a convolution module. The activation function in the feed forward module is used Swish activation and the intermediate hidden size is four times of the hidden size. The number of heads in the multi-head self-attention is 1/64 of the hidden size and relative positional embedding is used as positional encoding. Dropouts are applied similarly to the original model.

### 2.2. BEST-RQ Framework

In the pre-training step, BEST-RQ framework is used for the encoder. By applying vector quantization, BEST-RQ allows a task that masks a certain time of inputs with probability and predicts masked values like MLM. Inputs are normalized to mean 0, standard deviation1 and the masked inputs are converted noise sampled from mean 0, standard deviation 0.1.

The random projection and the random codebook are initialized with Xavier initialization and standard normal distribution. These parameters are fixed during training. A multiclass classification loss is used as the loss function instead of a contrastive loss by applying a random projection quantizer. Thanks to this, BEST-RQ enables simpler pre-training compared to wav2veac2 and has succussed to reduce the difference between pre-training and downstream tasks.

### 2.3. Temporal Gate Pooling

We designed a pooling method suitable for speaker identification by adapting a gate mechanism inspired by gmlp [11, 12] (Figure 2). TGP generates a gate in units of time from the hidden states and multiplies them with the hidden states to enable effective pooling. TGP generates the gate through a time-wise neural network (NN) and has a simple yet powerful learning ability. In addition, TGP supports multi-head process.

Consider hidden states $H = [h_1, h_2 \cdots h_N]$ with $h_t \in \mathbb{R}^d$. The hidden states are converted filter $F \in \mathbb{R}^{N \times d}$ and value $V \in \mathbb{R}^{N \times d}$ through two pointwise NNs:

$$F = HW_F + b_F, V = HW_V + b_V \quad (1)$$

The filter is conducted cross-token interactions through a timewise NN:

$$F' = FW_T + b_T \quad (2)$$

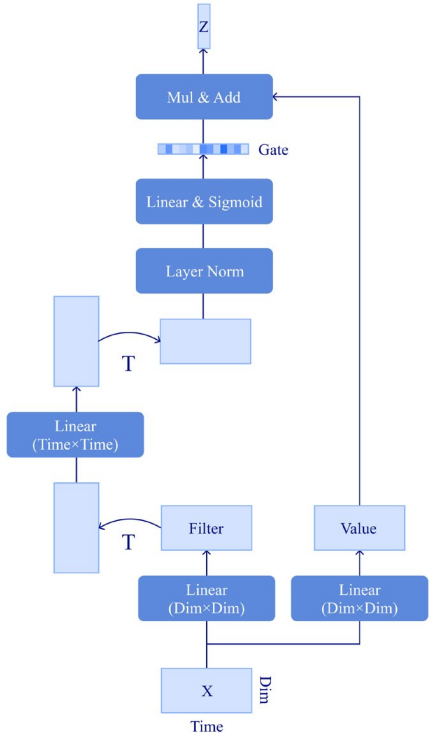

Figure 2: *Temporal Gate Pooling architecture. It supports multi-head processing.*

The filter is converted gate $G \in \mathbb{R}^{N \times 1}$ through a layer normalization, a regression NN and sigmoid:

$$G = Sigmoid(Layernorm(F')W_G + b_G) \quad (3)$$

Finally, the value and the gate are multiplied and summed along the time dimension to produce embedding $E \in \mathbb{R}^d$:

$$E = \sum_N V \odot G$$

where $\{W_F \in \mathbb{R}^{d \times d}, W_V \in \mathbb{R}^{d \times d}, W_V \in \mathbb{R}^d\}$ refer to weights of point-wise NNs, $\{b_F \in \mathbb{R}^d, b_V \in \mathbb{R}^d, b_V \in \mathbb{R}^d\}$ refer to biases of point-wise NNs. $W_F \in \mathbb{R}^{N \times N}$ and $b_F \in \mathbb{R}^N$ refer to the weight and the bias of the timewise NN and $b_F$ is initialized with 1. $\odot$ denotes elementwise multiplication.

## 3. EXPERIMENTS & DISCUSSION

### 3.1. Data

We adopted LibriSpeech dataset [13] which consists of 970 hours speech data as an unlabeled dataset for pre-training, and VoxCeleb1 dataset [7] for a speaker identification task. VoxCeleb1 dataset has 1251 speakers and 153516 utterances and is divided into 145265 of them for training and 8251 of them for testing as a speaker identification task.

Table 1: *Hyper-parameters for Conformer-based encoders.*

| Model | 256M | 512M | 768M | 256S |
|---|---|---|---|---|
| Hidden Size | 256 | 512 | 768 | 256 |
| FFN Hidden Size | 1024 | 2048 | 3076 | 1024 |
| Num Attn-Heads | 4 | 8 | 12 | 4 |
| Encoder Layers | 16 | 4 | 2 | 2 |
| Num Params (M) | 25.9 | 26.9 | 31.8 | 3.6 |
| Inference Time (ms) | 24.2 | 9.7 | 6.8 | 5.8 |

In this study, we converted raw data to a fixed length of 15 sec and extracted 80-channel log-mel spectrogram computed from a window of 25ms with a stride of 10 ms.

### 3.2. Pre-training

In the pre-training, we set encoders with Softmax to BEST-RQ. In this study, the mask length is set to 200ms and mask probability is set to 0.05. Also, the codebook contains 8192 vocabularies with 16 dimensions. Because the input length is reduced to 1/4 through the encoders, one vector in the BEST-RQ codebook contains 4 lengths of the input (40ms). The training used AdamW and linear-warmup-and-cosine-decay scheduler. The learning rate is 1e-4 and the number of warmup steps is 10000. Finally, the models learned 0.44M steps in a batch of 128.

### 3.3. Effective Hyper-Parameters

Four distinct models, 256M, 512M, 768M and 256S, with varying hidden sizes and numbers of layers was prepared (Table 1). Models 256M, 512M and 768M possess approximately equivalent number of parameters (num params), while model 256S exhibits a lower num params compared to the other three. These four models underwent pre-training prior to experimentation. Also, a model, 512M-B, without pre-training was prepared to compare with and without pre-learning. The assessment involved a performance comparison of the models based on Accuracy, utilizing speaker identification of VoxCeleb1. For training, an AdamW optimizer with a learning rate of 1e-4 and a linear-warmup-and-cosine-decay scheduler with 10,000 steps were employed, conducting training across a maximum of 80 epochs with a batch size of 64.

The graph (Figure 3) show that 256M exhibited the highest degree of precision, followed by 512M, 768M, 512M-B, and 256S, in that order. This result signifies that precision is fundamentally determined by their num params, with a particular emphasis on enhancing precision through an increase in the number of layers. However, it is noteworthy that 256M necessitates over twice the inference time when compared to 512M. Thus, it is concluded that 512M strikes the most effective balance between inference time and precision. Furthermore, the comparison between 512M and 512M-B underscores the efficacy of BEST-RQ for speaker identification.

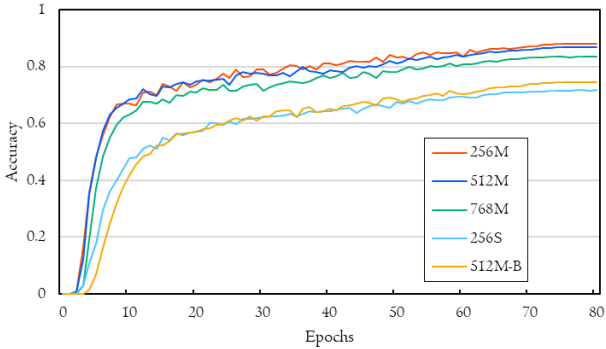

Figure 3: *Comparison of model learning progress.*

### 3.4. Pooling Methods

We conducted a comparative analysis between TGP and several statistical pooling methods and attentive pooling, including the self-attention pooling [15, 16, 17]. Mean pooling, mean and standard deviation pooling, max pooling and random pooling are included the statical pooling. For the encoder, we employed 512M Conformer as in Section 3.2. The evaluation metric remained consistent with Section 3.2, focusing on the accuracy of speaker identification. The training process involved employing an AdamW optimizer with a learning rate of 1e-4, accompanied by a linear-warmup-and-cosine-decay scheduler spanning 10,000 steps. The training was carried out with a batch size of 64, over a maximum of 120 epochs.

TGP demonstrated a superior accuracy of up to 4.0% compared to other existing pooling methods, thus establishing its efficacy (Table 2). Furthermore, adapting multi-head, the model achieved enhancement of 0.11% in accuracy compared single-head. Also, mean pooling, despite its simplicity, exhibited superior accuracy compared to self-attention pooling has learning ability. This result suggests the possibility of eliminating the need for complex pooling by improving the performance of the encoder.

### 3.5. Performance of the proposed method

We conducted a comparative analysis between several conventional models and the proposed model based on their num params and the accuracy of speaker identification of VoxCeleb1. The models subjected to comparison in this study, namely wav2vec2 and HuBERT [14], both adapt Transformer as encoders, employing distinct pre-training. Also, wav2vec2 takes raw, untreated audio data as input. HuBERT utilizes MFCC as its input representation. The comparative model data has been referred to from SUPERB benchmark [6].

Table 2: *The performance of different pooling methods for speaker identification.*

| Pooling | Accuracy on SID (%) |
|---|---|
| mean | 85.69 |
| mean & std | 84.92 |
| max | 83.82 |
| random | 83.18 |
| self-attention | 85.26 |
| self-attention (multi-head) | 86.44 |
| temporal gate | 87.02 |
| temporal gate (multi-head) | **87.13** |

Table 3: *Number of parameters and the performances of speaker identification models. For the lines with * notation, we referred the value from SUPERB benchmark [6].*

| Method | Params (M) | SID (%) |
|---|---|---|
| wav2vec2.0 Base* [2] | 95.04 | 75.18 |
| wav2vec2.0 Large* [2] | 317.68 | 86.14 |
| HuBERT Base* [14] | 94.68 | 81.42 |
| HuBERT Large* [14] | 316.61 | 90.33 |
| Ours | 28.51 | 87.13 |

Table 3 shows that HuBERT Large achieved the highest accuracy and the proposed method achieved the second-highest accuracy. Furthermore, the proposed method showcases a remarkably elevated level of accuracy, comparable to wav2vec2 Large with num params exceeding tenfold.

### 4. CONCLUSION

We investigated the impact of model hyper-parameters on the accuracy of speaker identification task. We showed the significance of balancing hidden size and the number of layers to achieve a harmony between accuracy and inference speed. Furthermore, we proposed a pooling method, Temporal Gate Pooling, and showed that the pooling method is effective for speaker identification task. Finally, our proposed method succeeded in achieving comparable performance to models with over tenfold num params.

### 5. FUTURE WORK

Future work in this field is anticipated to evolve with the emergence of more advanced contextual models and learning method. Additionally, as model precision advances, the role demanded of pooling might diminish, possibly rendering simpler pooling methods such as mean pooling sufficient.